# Brain Tumor Detection using Convolutional Neural Networks with Skip Connections

Aupam Hamran, Marzieh Vaeztourshizi, Amirhossein Esmaili, Massoud Pedram

Dept. of Electrical and Computer Engineering

University of Southern California

*Abstract*—In this paper, we present different architectures of Convolutional Neural Networks (CNN) to analyze and classify the brain tumors into benign and malignant types using the Magnetic Resonance Imaging (MRI) technique. Different CNN architecture optimization techniques such as widening and deepening of the network and adding skip connections are applied to improve the accuracy of the network. Results show that a subset of these techniques can judiciously be used to outperform a baseline CNN model used for the same purpose.

I. INTRODUCTION

The brain is one of the most complicated organs inside the human body that works with a very large number of cells. Brain tumors increase when there is an unregulated division of cells that forms an irregular mass. This group of cells will affect the normal function and activity pattern of the brain and damage the healthy brain cells [1]. X-ray images are typically used for evaluating and recognizing the body's tumor growth. Detecting a brain tumor is also possible with medical examination imaging techniques such as Magnetic Resonance Imaging (MRI) and computed tomography (CT) [2]. Digital image processing is a critical task in analyzing the MRI scans.

Tumor development takes place inside the skull. Tumors cause extreme brain pressure, which spread across the entire brain region. Some of these tumors can be malignant and lead to Cancer, a major cause of death and responsible for about 13 percent of all deaths worldwide. Today, a radiologist identifies the brain tumors by visual inspection. Tumor classification process can be extremely time consuming whose accuracy is a function of skills and experience of the radiologist. With the increase in the number of patients with MRI scans, the amount of data to be analyzed daily is large, which makes the reading based on visual interpretation expensive and slow.

Classification of brain tumors into various pathological types is more challenging than the binary classification. The related challenges are attributed to factors such as high variations in shape, size, and intensity of the same tumor type [3] and similar appearances for different pathological types [4]. A false negative diagnosis of a brain tumor can lead to a decreased chance of survival for the patient. A false positive diagnosis can result in traumatic experiences for the patients and their families. To overcome drawbacks of the manual diagnosis, there has been a surge of interest in designing automated image processing systems [5][6][7]. Many researchers have suggested techniques to improve the quality of the computer aided design (CAD) system that classifies tumors in brain MRI.

Machine learning (ML) methods used in the classification process are usually based on different steps such as preprocessing, dimension reduction, feature extraction, feature selection, and classification. The feature extraction represents the crucial phase in an effective CAD system [8]. It is a challenging task and requires prior knowledge about the problem domain since the classification accuracy depends strongly on the quality of the feature extraction. Traditional techniques for feature extraction can be classified into three types: spatial domain features, wavelet and frequency features and contextual and hybrid features. The state-of-the-art CAD methods yield an improved performance due to the use of deep learning (DL) technology.

DL represents a subset of ML, which does not require handcrafted features [9][10]. It has been successfully employed to reduce the gap between human brain and computers in many applications, including pattern recognition, object detection, natural language understanding, sensory data perception, and so on. DL has already surpassed state-of-the-art methods in several fields such as generating text [11], face verification [12], image description [13], the game of Go [14], and grand challenges [15]. According to [16] there were about 220 paper publication focusing on DL for medical image diagnosis in 2019. Around 190 of these works used Convolutional Neural Network (CNN). DL permits the use of pre-trained CNN models such as AlexNet [17], GoogLeNet [18], ResNet-34 [19] for medical image diagnosis especially for brain tumor classification.

CNNs have exhibited high performance as an inference engine after being carefully trained on huge, labeled datasets such as ImageNet [15] which contains more than one million images. However, it is hard to exploit such deep CNNs in the medical field. First, the size of the medical datasets is generally small because such datasets need the availability of expert radiologists to manually examine and label the images, which is time-consuming, laborious, and costly. Second, training deep CNNs is a complex process especially with a small dataset. This is because of the pitfalls for over-fitting or divergence. Third, domain expertise is needed to repeatedly revise the model and adjust the learning parameters of DL models to achieve better performance.

This paper presents various neural network models for brain tumor classification. The remainder of the paper is organized as follows. In Section II, related works on brain tumor detection of MRI images using CNNs are reviewed. A review of some of the main topics discussed in the paper is presented in Section III and the details of the proposed framework are discussed in Section IV. In Section V, the effectiveness of the proposed framework is evaluated for brain tumor datasets and the paper is concluded in Section VI.

## II. RELATED WORK

In this section, some of the relevant prior works related to automated brain tumor diagnosis using CNNs are reviewed. This review is not meant to be comprehensive. In fact, we have only considered some of the recent related works that have utilized the same brain MRI dataset as us -- see [20] and [21].

In [22], a CNN was utilized to classify the brain MRI images to flag the presence or absence of brain tumors in the image. Three test cases were considered, where the convolution kernel size (size 2×2 or 3×3) and max pooling size (*e.g.*, 2×2) and hyper parameter values for training were different for each test case. Each of these cases were trained with 5 different epoch values {10, 20, 30, 40 and 50}. The number of convolutional (and max pooling) layers were considered as 2, 4 and 5 with two fully connected (FC) layers (with similar neuron counts). To check the robustness of the proposed network, the training data and the testing data were selected from [20] and [21], respectively.

In [23], first the MRI dataset (*e.g.,* from [21]) was pre-processed (converted from RGB to grayscale and scaled down to 224×224 pixels). Next, a CNN model was proposed consisting of six convolutional layers with different kernel sizes (1×1, 3×3, and 5×5, each followed by a max pooling layer with size 2×2). The next layers were two FC considering 50% dropout. Batch size of 32 was considered when training the network for 30 epochs.

A CNN model with three 3×3 convolution layers (each followed by a max pooling layer) and two FC layers was trained for 250 epochs in [24] on the [21] dataset where the pixel values of the original images were scaled down. The batch size was set at 32. Next, data augmentation with random rotation, flip and zoom was applied to the image and the augmented images were added to the original dataset and fed to the model for retraining with the same parameters. It was shown that data augmentation provided better network accuracy. Due to the small size of brain tumor datasets generally, [25] proposed to use transfer learning [26]. In practice, several pre-trained CNN models with ImageNet [15] dataset were utilized as the backbone of the DL. A data augmentation technique consisting of rotation, horizontal and vertical flip was also performed on the normalized and scaled MRI images (*e.g.,* [21]). The different CNN models of ResNet, GoogleNet and EfficientNet (initialized with weights obtained by ImageNet [15] dataset) were considered with EfficientNet outperforming the other networks.

## III. BACKGROUND

In this section, a review of some important concepts related to CNNs and skip connections are provided. A brief discussion of MRI images and their usage in brain tumor detection concludes this section.

### A. CNNs

A CNN is a feedforward neural network (NN) architecture from the class of DL networks which was inspired by the human Visual Cortex [27]. These networks are useful for processing data with grid-like topologies such as pixel data processing and recognizing patterns in images [28]. CNNs are highly applicable in the fields of image processing, face recognition, object detections, computer vision and natural language processing.

Each CNN model contains different types of layers. The input of each layer in a CNN is called the input feature map while the output is called the output feature map. The most important layer in a CNN, which is also always the first one, is the convolutional (conv) layer. A conv layer is responsible for extracting features from the data. It consists of a set of 3D filters. During training of the network, each filter can be considered as a set of trainable cube-shaped weights. A filter can be considered as a set of 2-D kernels.

For each filter, the depth (called the number of channels) of the filter, which is also the number of kernels, is always the same as the depth of the input feature map. When designing a CNN architecture, the number of filters for each conv layer and the size of the kernels (*i.e.*, height and width of the filter) are parameters chosen by the designer. Note that depending on these parameters, the conv layer can contain the most user-specified parameters of the network [29].

The pooling layer, known for down-sampling, is used to lower the size of the convolved output feature map. Similar to the conv layer, in the polling operation, a weight-less filter is applied to the input feature map and depending on the type of the pooling operation, the values of output feature map are calculated. As an example, in max pooling, the maximum of the values of the input feature map covered within each kernel of the filter represents the corresponding output feature map in max pooling operation. For average pooling, the average of the values within each kernel of the filter represent the corresponding output feature map [29].

Each CNN model can contain one or more conv layers followed by pooling layers. The final layers of the CNN architecture consist of fully connected (FC) layers. Each of the mentioned layers can have different parameters (*i.e.*, weights and biases) that need to be optimized during the training of the network to perform different tasks on the input data.

## B. Skip Connections in NNs

With larger (deeper) networks, to solve the accuracy degradation issue, reference [30] proposed to use skip connections. The goal of a skip connection is to add the output of a layer in the network (*i.e.,* identity mapping) to the output of a deeper layer before applying the activation function of the deeper layer. Fig. 1 shows an example of a basic block of skip connections, which we call a skip connection block (SCB). The number of conv layers in the direct path of the SCB may be varied. In addition, a set of conv layers may also be applied to the skipped value $X$ in Fig. 1 on the upper path before the addition. The added skip connections do not add many extra parameters or increase the computational complexity of the network [30].

## C. Brain Tumor Detection

A brain tumor is an appearance of an unwanted mass of tissue in the brain due to the abnormal growth of brain cells which can be malignant (cancerous) or benign (non-cancerous). According to the National Brain Tumor Society [31], an estimated 700,000 people are living with a brain tumor in the United States where around 1800 will die from a malignant tumor. Therefore, early diagnosis of this deadly disease is highly critical for suitable treatments.

MRI is a form of scan that produces a detailed image of the inside of the tissue using high magnetic fields and radio waves [32]. A benign tumor has distinct, smooth and regular borders while a malignant tumor has irregular borders and grows faster than a benign tumor. Fig. 2 shows an MRI image of a non-cancerous and a cancerous brain tumor. To manually detect a brain tumor, an experienced specialist doctor is required to analyze the tumor which can be a time consuming process assuming the availability of the expert in the location. Therefore, automating the detection of brain tumors in their early stages is highly encouraged. In the early stages of research, different image processing and data mining techniques have been employed to analyze brain tumors [2]. Recently, with the advent of ML techniques, these networks have been employed in the process of tumor diagnosis with great success.

## IV. NETWORK ARCHITECTURE OPTIMIZATION

In this work, we describe our investigation to assess the efficacy of different network architectures targeting brain tumor detection. The objective of this study was to optimize the inference accuracy of the network for the classification of brain tumors to benign and malignant types. We considered different architectures for CNNs as well as the combination of CNNs and SCBs.

To obtain different CNN architectures, we used network widening (*i.e.*, changing the number of filters in a conv layer) and network deepening (*i.e.*, adding more conv + pooling layers to the network) techniques. We also changed the number of filters in the conv layer of the SCB. Note that in the networks, batch normalization (BN) and max pooling (Pool) are applied after each convolutional layer. We call the combination of a conv layer, BN, and Pool a conv block (CB).

The baseline architecture, which is depicted in Fig. 3, had two convolutional layers with sizes of 32 and 64 filters and two fully connected layers with sizes of 1024 and 256 neurons, respectively (this is denoted as Arch-1). Since the problem is a binary classification one, the last FC layer has one neuron for its output. In all networks considered in this work, the kernel sizes of the conv and pooling layers are set as 3×3 and 2×2, respectively. In this study, we assumed the size of the fully connected layer remains unchanged. For any architectures proposed, the impact of data augmentation on the output accuracy was also assessed. For data augmentation, we added horizontally and vertically flipped images (randomly generated) from the training dataset to the original set to obtain a larger (augmented) dataset.

To generate new network architectures, first we doubled the number of filters in both convolutional layers of the base model to obtain Arch-2 (network widening). Arch-3 was obtained by adding a new layer to the base model (network deepening) while in Arch-4, widening was performed on Arch-3 with doubled filter counts for all the convolutional layers (network deepening and widening of the original layers). Arch-5 was our largest model with four convolutional layers (deepening).

When evaluating the impact of incorporating SCBs into the CNN, first, Arch-6 (depicted in Fig. 4) was defined as the base model with a single SCB. In this study, similar to Fig. 1, on the direct path of the SCB, two conv layers were considered with a single conv layer located on the upper path for the skip connection. For all the conv layers in the SCB, the number of filters was chosen to be the same. This number represented by the SCBS parameter (*e.g.*, SCBS = 3 in Fig. 4).

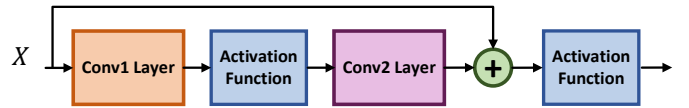

**Fig. 1.** An example of a skip connection block [30].

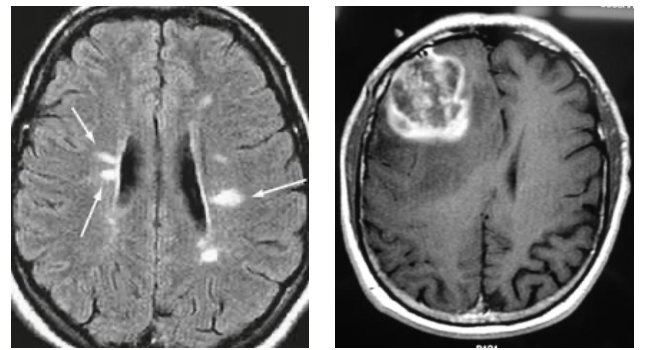

**Fig. 2.** An example of an MRI image of a (a) benign and (b) malignant brain tumor taken from [21].

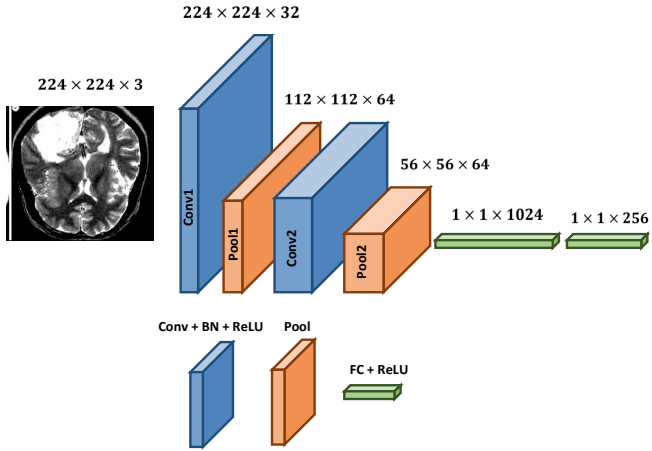

**Fig. 3. The baseline architecture (Arch-1), which is the smallest CNN utilized for classification of brain tumors.**

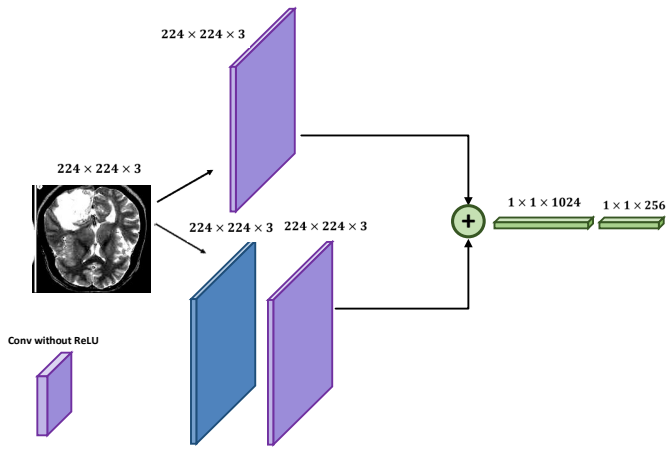

**Fig. 4. The baseline architecture (Arch-6) with SCB.**

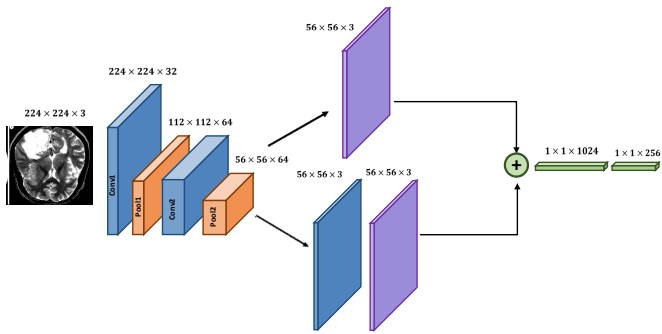

**Fig. 5. Arch-7 as an example of a network which combines CNN and SCB.**

In this work, the SCB in the network architecture is always located after the CB, which means the output of the CB (*e.g.*, the output of Pool2 layer in Fig. 3) is connected to the input of the skip connection (*i.e.*, $X$ in Fig. 1). The CNN baseline model was combined with SCBS = 3 in Arch-7 and represented in Fig. 5 as an example for the networks which combined CB and SCB. In Arch-8, the baseline model was deepened (to four layers like Arch-5) whereas in Arch-9, the network was both widened and deepened (to three layers like in Arch-4.) Finally, SCBS was increased to eight to see the impact of number of SCB filters on the network accuracy for the CNN baseline model combined with SCB.

## V. RESULTS AND DISCUSSIONS

We begin this section by explaining the experimental setup considered in this work. Next, we present the network accuracy values obtained for the different architectures with and without the data augmentation technique.

### A. Experimental Setup

Results of the experiments mentioned in Section IV are presented in Table 1 and Table 2. Table 1 presents the results for different configurations of CNN-based networks without SCB while Table 2 presents different configurations for CNN-based networks with SCB. For the dataset, we have considered the two datasets of [20] with 253 data values (155 cancerous and 98 non-cancerous) and [21] with 3000 data values (1500 cancerous and 1500 non-cancerous). Similar to [22], since training accurate DL models requires a large number of images, we have trained the network with the larger dataset [21] and tested that on the smaller dataset [20].

For the data augmentation technique, the mentioned flipping techniques are applied to all the samples in [21] and added to the original set. The results will be provided in the next section. For all networks, we set the training batch size to 15 and the number of epochs to 60.

### B. Discussion of the Results

In this section, the network performance, which is the inference accuracy of the different networks, is reported and discussed.

#### 1) Data Augmentation

Based on the results, in most cases (seven out of ten architectures), we see no improvement in neural network's output accuracy using augmentation. In the case of CNN models without SCB, for one case of widening (Arch-2) and one case of deepening (Arch-5), we see that augmentation has increased the network accuracy by on average 0.6%. The highest accuracy achieved in the case of CNN models is also obtained with augmentation applied to the largest network proposed (*i.e.*, Arch-5).

When considering the combination of CNN and SCB, the network accuracy is increased using data augmentation in only one case of deepening for Arch-8 (which is similar to Arch-5 in the CNN part). As it can be seen from the results, the impact of data augmentation for the utilized training dataset of [21] is dependent on the architecture of the network.

**Table 1.** Accuracy values for various neural network models

|        | AUGMENTATION | Convs |       |       |       | FCs  |     | ACCURACY |
|--------|--------------|-------|-------|-------|-------|------|-----|----------|
|        |              | conv1 | conv2 | conv3 | conv4 | FC1  | FC2 |          |
| Arch-1 | No           | 32    | 64    | -     | -     | 1024 | 256 | 99.20    |
|        | Yes          | 32    | 64    | -     | -     | 1024 | 256 | 99.20    |
| Arch-2 | No           | 64    | 128   | -     | -     | 1024 | 256 | 98.80    |
|        | Yes          | 64    | 128   | -     | -     | 1024 | 256 | 99.20    |
| Arch-3 | No           | 32    | 64    | 128   | -     | 1024 | 256 | 99.20    |
|        | Yes          | 32    | 64    | 128   | -     | 1024 | 256 | 98.80    |
| Arch-4 | No           | 64    | 128   | 256   | -     | 1024 | 256 | 99.20    |
|        | Yes          | 64    | 128   | 256   | -     | 1024 | 256 | 99.20    |
| Arch-5 | No           | 32    | 64    | 128   | 256   | 1024 | 256 | 98.80    |
|        | Yes          | 32    | 64    | 128   | 256   | 1024 | 256 | **99.60** |

**Table 2.** The accuracy values for CNNs with Skip Connections

|         | AUGMENTATION | Convs |       |       |       | SCBs  | FCs  |     | ACCURACY |
|---------|--------------|-------|-------|-------|-------|-------|------|-----|----------|
|         |              | conv1 | conv2 | conv3 | conv4 | conv1 | FC1  | FC2 |          |
| Arch-6  | No           | -     | -     | -     | -     | 3     | 1024 | 256 | **99.60** |
|         | Yes          | -     | -     | -     | -     | 3     | 1024 | 256 | 98.80    |
| Arch-7  | No           | 32    | 64    | -     | -     | 3     | 1024 | 256 | **99.60** |
|         | Yes          | 32    | 64    | -     | -     | 3     | 1024 | 256 | 98.80    |
| Arch-8  | No           | 32    | 64    | 128   | 256   | 3     | 1024 | 256 | 98.40    |
|         | Yes          | 32    | 64    | 128   | 256   | 3     | 1024 | 256 | 98.80    |
| Arch-9  | No           | 64    | 128   | 256   | -     | 3     | 1024 | 256 | **99.60** |
|         | Yes          | 64    | 128   | 256   | -     | 3     | 1024 | 256 | 98.80    |
| Arch-10 | No           | 32    | 64    | -     | -     | 8     | 1024 | 256 | **99.60** |
|         | Yes          | 32    | 64    | -     | -     | 8     | 1024 | 256 | 99.20    |

### 2) Network Architecture

Evaluating the accuracy values of the different network architectures, it can be seen that in the case of CNNs without SCBs (see Table 1), widening the network by one layer did not change the accuracy of the network (Arch-1 vs Arch-3 without data augmentation). The same result also holds when a deepened network (Arch-2) was widened (Arch-4). However, a combination of the proposed techniques (deepening, widening, and augmentation) produced the best accuracy result in Arch-5 where 99.60 accuracy was achieved.

On the other hand, for networks with SCBs, in almost all architectures (except Arch-8), the best accuracy value (*i.e.*, 99.6%) was achieved without any data augmentation. With the same architectures, sometimes adding SCBs resulted in higher accuracy (Arch-1 vs Arch-7) while sometimes the CNN alone resulted in better accuracy (Arch-5 vs Arch-8). Increasing the number of filters in the SCB (*i.e.*, SCBS) only had an impact when the data augmentation technique was utilized (Arch-7 vs Arch-10). Note that in this study, we did not change the size and number of FC layers. The reason was that we wanted to assess the efficacy of the convolutional and skip connection layers on the network accuracy. In fact, we have increased the number of FC layers for Arch-5 (best network in the CNN models) and observed that the network accuracy was lower than the maximum values obtained (*i.e.*, 99.6%).

## VI. CONCLUSIONS

In this paper, we studied the efficacy of Convolutional Neural Networks (CNNs) with the addition of skip connections to detect brain tumors from Magnetic Resonance Imaging (MRI) images. Considering different architectural CNNs such as widening and deepening techniques and the position of the skip connection, the accuracy of the network was compared with a baseline CNN model. The results showed that techniques such as deepening, widening, augmentation, and utilizing SCBs improved the accuracy results in general and the best results were obtained when a subset of the techniques were employed judiciously.